\documentclass[letterpaper,nameyear,ecta,12pt]{article}
\usepackage{amsmath}
\usepackage{pgfplots}
\usepackage{amsthm,amssymb,mathtools}
\usepackage{enumitem}
\usepackage[linesnumbered,ruled,vlined]{algorithm2e}
\usepackage{comment,todonotes}
\usepackage{natbib}
\usepackage[colorlinks,citecolor=blue,linkcolor=blue]{hyperref}
\usepackage{setspace}
\usepackage[margin=1.25in]{geometry}

\theoremstyle{definition}
{}
\newtheorem{definition}{Definition}{}
\newtheorem{lemma}{Lemma}{}
\newtheorem{proposition}{Proposition}{}
\newtheorem{theorem}{Theorem}{}
{}
\newtheorem{example}{Example}{}
{}
\newtheorem{remark}{Remark}{}
\newtheorem{claim}{Claim}{}

\title{ Core nonemptiness for $2/3$-convex games\thanks{We are grateful to the participants of European Meeting on Game Theory 2026 (SING21) 
and Asian Game Theory Conference 2026 (AGT2026) 
for their helpful comments and suggestions.  This work was partially supported by JSPS KAKENHI Grant Numbers  23K01312, 26K21170, and 26K04829 and JST ERATO Grant Number JPMJER2301, Japan.}}

\author{Keisuke Bando\thanks{Department of Industrial and Systems Engineering, Faculty of Science and Technology, Keio University, email: bando@ae.keio.ac.jp.} \and Toshiyuki Hirai\thanks{Department of Economics, Hosei University, e-mail: hirai@hosei.ac.jp.} \and Kenzo Imamura\thanks{Department of Economics, University of Tokyo, email: imamurak@e.u-tokyo.ac.jp.} 
\and Ryuhei Mizutani\thanks{Faculty of Engineering, Tokyo University of Agriculture and Technology, email: \mbox{mizutani}@go.tuat.ac.jp}
}

\begin{document}

\maketitle
\onehalfspacing

\begin{abstract}
We study the core of $2/3$-convex games, a class of
transferable-utility (TU) games that generalizes convex games,
which requires at least two of any three distinct coalitions satisfy the supermodular inequality.
We show that every $2/3$-convex game 
admits a marginal worth vector in the core, and hence has a
nonempty core. Our proof is constructive and
 yields a linear-time algorithm for finding such a core allocation.
%requires only
%{\color{magenta}$O(n)$ evaluations of the characteristic function.}
We also show that $2/3$-convexity does not imply permutational convexity
in general, while the implication holds under an additional
superadditivity condition on pairs of singletons.
As an application, we introduce a TU model motivated by the OSCE’s consensus-minus-two procedure and show that the resulting game is \(2/3\)-convex but need not be convex.
\end{abstract}

\section{Introduction}\label{Sect:introduction}

Convexity is a fundamental sufficient condition for core nonemptiness
in transferable-utility (TU) games. In a convex game, the marginal worth
vector associated with every ordering of the players belongs to the
core \citep{Shapley1971Cores}.\footnote{
\citet{Ichiishi1981} establishes the converse: if every marginal
worth vector belongs to the core, then the game is convex.}
This property connects a global stability requirement---that every
coalition receive a total payoff at least as large as its worth---to
a simple sequential allocation procedure. A natural question is how
much convexity can be relaxed while retaining at least one core
allocation of this form.

We study a relaxation of convexity, called \emph{$2/3$-convexity},
in the framework of TU games. This notion is the supermodular
counterpart of $2/3$-submodularity, introduced by
\citet{MizutaniYoshida2024,MizutaniYoshida2026TwoThirdsSubmodular}
in combinatorial optimization.\footnote{
\citet{MizutaniYoshida2024,MizutaniYoshida2026TwoThirdsSubmodular}
develop polynomial-time algorithms for minimizing
$2/3$-submodular functions.}
A TU game is $2/3$-convex if at least two of any three distinct coalitions satisfy the supermodular inequality.

Our main result shows that every $2/3$-convex game 
admits at least one marginal worth vector in the core, and thus
has a nonempty core.
We provide a technical lemma for the larger  class of games,  called \emph{$3$-distant convex
games,}  originally introduced by \citet{Mizutani2024KDist}.
For these games, we characterize when a given marginal worth vector
belongs to the core in terms of a simple local condition
(Lemma~\ref{thm:mw}).
Using this characterization, we construct a marginal worth vector in the core of 2/3-convex games.
The construction requires only $O(n)$ evaluations of the
characteristic function  and $O(n)$ arithmetic operations, where $n$ is the number of players.

We then compare $2/3$-convexity with permutational convexity, introduced
by \citet{Granot1982Perm}.\footnote{
Several relaxed notions of convexity is introduced for TU games, including semiconvexity \citet{DriessenTijs1985Semiconvex}, average convexity \citep{Sprumont1990PMAS, IU1993IJGT}, almost convexity \citep{Nunez1998almostconvex},
and hyper-additivity \citep{Bahel2021Hyperadditive}.
\citet{Driessen1986KConvex} and  \citet{Bahel2026Preconvex} also introduce a class of TU games parameterized by a positive integer $k$, 
called $k$-convex games and $k$-preconvex games, respecitvely, which coincide with convex games when $k = n$.
} Permutational convexity also guarantees
that at least one marginal worth vector belongs to the core.\footnote{ 
\citet{Alidaee1991} shows that the converse does not hold:
the existence of a marginal worth vector in the core does not
imply permutational convexity.}
We show that a $2/3$-convex game need not be permutationally convex (Example \ref{ex:non-perm}).
Thus, our main result does
not follow from this existing sufficient condition.
Finally, we show that $2/3$-convexity implies permutational
convexity under superadditivity on pairs of singletons.
%the worth of each two-player coalition is at least the sum
%of its members' singleton worths.

We also provide an institutional application  motivated by the OSCE's
consensus-minus-two procedure. Under this procedure, neither of the two
States involved in a dispute can veto a decision directing them to seek
conciliation
\citep[Annex~4, paras.~1 and 4]{CSCE1992Stockholm};
see also \citet{OSCEDecisionMaking}.
We model the value generated by such a decision as a TU game and show that the resulting game is \(2/3\)-convex but need not be convex.
% We consider a TU game in which the characteristic function represents the value generated by the adopted decision.
% We show that the resulting game is \(2/3\)-convex, although it need not
% be convex.

The remainder of the paper is organized as follows.
Section~2 introduces $2/3$-convex games and provides an
illustrative example.
Section~3 presents and proves the main results and examines
the relationship with permutational convexity.
Section~4 provides an application. 
Section~5 concludes.

\section{Model}
Let $N = \{1, \ldots, n\}$ be a finite set of players.
Throughout the paper, we assume that $n \ge 3$.
A nonempty subset of $N$ is called a \emph{coalition}.
A real-valued function $v$ on $2^N$ is called a \emph{characteristic function}, where $v(S)$ denotes the worth of $S \subseteq N$ and $v(\emptyset) = 0$.
The pair $(N, v)$ is called a \emph{transferable-utility} (TU) game.
%{\color{orange} Throughout the paper, we assume that 
%$v(N) \ge \sum_{i\in N}v(\{i\})$.\footnote{This is a very mild condition as it is weaker than essentiality, that is, $v(N)>\sum _{i\in N}v(\{ i\} )$, which is usually assumed.}}

A vector $x \in \mathbb{R}^N$ is called an \emph{allocation}.
For each allocation $x$ and $S \subseteq N$, we denote
$x(S) = \sum_{i \in S} x_i$ with $x(\emptyset) = 0$.
The \emph{core} of $(N, v)$ is defined as the set of allocations $x$
such that $x(N) = v(N)$ and
$x(S) \ge v(S)$ for all $S \subseteq N$.
In this case, $x$ is called a \emph{core allocation}.

For each $S, T \subseteq N$, let
$
D(S, T) = v(S \cup T) + v(S \cap T) - ( v(S) + v(T) ).
$
We say that $(N, v)$ is a \emph{convex game} if 
$D(S, T) \ge 0$  for all $S, T \subseteq N$.
Given any ordering of players $i_1, i_2, \ldots, i_n$,
the corresponding \emph{marginal worth vector} is defined by
\[
m_{i_\ell} = v(\{i_1, \ldots, i_\ell\}) - v(\{i_1, \ldots, i_{\ell -1}\})
\quad \text{for all } \ell = 1, \ldots, n.
\]
\citet{Shapley1971Cores} shows that every marginal worth vector is in the core,
and hence the core is nonempty for every convex game.

We consider a relaxed notion of convexity introduced by \citet{MizutaniYoshida2024, MizutaniYoshida2026TwoThirdsSubmodular}.
\begin{definition}\label{def:relaxations}
A game $(N,v)$ is \emph{$2/3$-convex} if, for any three distinct
coalitions $S_1,S_2,S_3\subseteq N$, at least two of
$D(S_1,S_2)$, $D(S_1,S_3)$, and $D(S_2,S_3)$ are nonnegative.
\end{definition}
 We often say that a characteristic function $v$ is $2/3$-convex if it satisfies the condition in Definition \ref{def:relaxations}.

The following example illustrates a $2/3$-convex game that may not be convex.

% {\color{red}[The following example is a simpler version of two air-port games.]}

\begin{example}
Let $(N,v)$ be a convex game with $N=\{1,2,3\}$, and let $c>0$.
Define
\[
v'(S)=
\begin{cases}
v(S)-c, & \text{if } S=\{1,2\},\\
v(S),   & \text{otherwise}.
\end{cases}
\]
The cost $c>0$ is an unavoidable coordination cost incurred when
players 1 and 2 carry out a joint project without player 3. We
interpret coalition values as the net values generated conditional
on carrying out the project. For example, players 1 and 2 may be a
design firm and a manufacturer using incompatible technical
standards, while player 3 is a systems integrator whose participation
eliminates the additional costs of redesigning interfaces and
correcting communication errors.
% The cost \(c>0\) is an additional transaction cost incurred when
% players \(1\) and \(2\) transact without player \(3\).
%  For example, players 1 and 2 may be a property seller and a buyer,
% respectively, while player 3 is a real-estate agent whose participation
% eliminates the additional costs of arranging and completing a direct
% transaction.

Then, $(N, v')$ may not be convex because 
$$
v'(\{1, 2\}) = v(\{1, 2\}) - c < v(\{1\}) + v(\{2\})  = v'(\{1\}) + v'(\{2\})
$$
holds
for sufficiently large $c > 0$.
On the other hand, only this pair violates the convexity condition.
This implies that $v'$ is 2/3-convex.

For a numerical illustration, let \(v\) be the unanimity game on
\(\{1,2\}\), that is,
\[
v(S)=
\begin{cases}
1, & \text{if }\{1,2\}\subseteq S,\\
0, & \text{otherwise},
\end{cases}
\]
and let \(c=4\). 

The marginal worth vectors associated with the orders \((1,2,3)\)
and \((1,3,2)\) are, respectively,
$(m_1, m_2, m_3) = (0,-3,4)$ and $(m_1, m_3, m_2) = (0,0,1)$.
The first vector does not belong to the core because its second
component is smaller than \(v'(\{2\})=0\), whereas the second vector
belongs to the core. Moreover, the Shapley value \citep{Shapley1953Value} of \(v'\) is
$\left(-\frac16,-\frac16,\frac43\right)$,
which does not belong to the core.
\end{example}

\section{Main result }

\subsection{Core nonemptiness}
Our main result establishes the core nonemptiness for $2/3$-convex game 
by showing that at least one marginal vector is in the core.
%in every $2/3$-convex game with at least three players.

\begin{theorem}\label{thm:core}
For every $2/3$-convex game, there exists a marginal worth vector that is in the core.
Therefore, the core is nonempty.
%Every $2/3$-convex game with $n\ge3$ admits a marginal worth vector
%in the core. In particular, its core is nonempty. Given an evaluation
%oracle for $v$, such a vector can be constructed using $O(n)$ oracle
%evaluations.
\end{theorem}

We first establish a technical lemma for a more general class of games.
A TU game
$(N, v)$ is called \emph{3-distant convex} if 
$v(S \cup T) + v(S \cap T) \ge v(S) + v(T)$ for all $S, T \subseteq N$ with $|S \triangle T| \ge 3$ where $S \triangle T = (S \setminus T) \cup (T \setminus S)$.
This concept was originally introduced by \citet{Mizutani2024KDist} in the context of
combinatorial optimization.

\begin{lemma}[\citet{Mizutani2024KDist}]\label{lem:2/3conveximplies3convex}
Every 2/3-convex game is $3$-distant convex.\footnote{
The converse does not hold.
For example, every $3$-person superadditive game is $3$-distant convex while they may not be 2/3-convex.}
\end{lemma}
% \citet{Mizutani2024KDist} introduced this concept shows in the context of
% combinatorial optimization, and showed that every $2/3$-convex game is $3$-distant convex for $3$-distant convex games.%\footnote{
% %{\color{red}The converse does not hold.
% %For example, every $3$-person superadditive game is $3$-distant convex while the }}

% %In a $3$-distant convex game,
% %no marginal vector is in the core even if the core is nonempty,
% %as illustrated in Example~\ref{ex:empty_core}.
% %We show that verifying whether a marginal vector belongs to the core is straightforward in 3-distance convex games.
% %Building on this result, we further provide a class of TU games in which at least one marginal worth vector lies in the core.

% %{\color{red}We consider a superadditive and $0$-nomalized game $(N,v)$.
% %In the following claim, we use the fact that a 2/3-convex game is a 3-distant %convex game [Check whether this assumption can be dropped.]}  
% %Superadditive is not necessary.  

The following result characterizes when a given marginal worth vector is in the core 
by a simple local condition for $3$-distant covex games.
\begin{lemma}\label{thm:mw}
Let $(N,v)$ be a $3$-distant convex game and $m$ be a marginal worth vector corresponding to an ordering of players $i_1,i_2,\ldots,i_n$. Then, 
%For any marginal worth vector $m$,
 $m$ is in the core if and only if there exists no  $k=0,1,\ldots,n-2$ such that 
% \begin{align}\label{k+1k+2}
% v(\{ i_1,...,i_k,i_{k+1})+v(\{ i_1,...,i_k,i_{k+2}\} )>v(\{ i_1,...,i_k,i_{k+1},i_{k+2}\} )+v(\{ i_1,...,i_k\} ).
% \end{align} 
\begin{align}\label{k+1k+2}
     v(\{ i_1, \ldots,i_k,i_{k+2}\} ) - v(\{ i_1,\ldots,i_k\} )>v(\{ i_1,\ldots,i_k,i_{k+1},i_{k+2}\} ) - v(\{ i_1,\ldots,i_k, i_{k+1}\} ),
\end{align}   
where $\{ i_1,...,i_0\} =\emptyset $.  
\end{lemma}

The right-hand side of the above equation represents the marginal contribution of 
$i_{k+2}$
 in the original order, whereas the left-hand side represents the marginal contribution 
 when $i_{k+2}$ skips $i_{k+1}$ and moves ahead of $i_{k+1}$.
 Therefore, this condition means that each player has no incentive to move one position forward in the order.

\bigskip
\noindent
{\bf Proof of Lemma \ref{thm:mw}.}
We first show the ``only if'' part.  %If $v(\{ i_1\} )+v(\{ i_2\} )>v(\{ i_1,i_2\} )$, then 
%$v(\{ i_2\} )>v(\{ i_1,i_2\} )-v(\{ i_1\} )=m_{i_2}$.  Thus, $m$ is not in the core.  
Assume that there exists some  $k=0,1,...,n-2$ such that
\begin{align*}
     v(\{ i_1,\ldots,i_k,i_{k+2} \}) - v(\{ i_1,\ldots,i_k\} )>v(\{ i_1,\ldots,i_k,i_{k+1},i_{k+2}\} ) - v(\{ i_1,\ldots,i_k, i_{k+1}\} ).
\end{align*} 
Let $m'$ be a marginal worth vector
corresponding to the ordering by swapping only $i_{k+1}, i_{k+2}$; that is,
$i_1,\ldots, i_{k+2}, i_{k+1},  \ldots, i_n$. 
Then, $m'_{i_\ell} = m_{i_{\ell }}$ for all $\ell = 1, \ldots, k$
and  $m'_{i_{k+2}} >  m_{i_{k+2}}$ by the above inequality.
Thus, $v(\{ i_1,\ldots,i_k,i_{k+2} \})
= \sum^{k}_{\ell = 1} m'_{i_{\ell }} + m'_{i_{k+2}} >
\sum^{k}_{\ell = 1} m_{i_{\ell }} + m_{i_{k+2}}$, where $\sum ^0_{\ell =1}m_{i_\ell }=\sum ^0_{\ell =1}m'_{i_{\ell }}=0$.
Thus, $m$ is not in the core.

We next show the ``if'' part.
Assume that $m$ is not in the core.  Then, there exists some coalition $S$ such that $\sum _{i\in S}m_i<v(S)$.  
Denote $S=\{ i_{\sigma (1)},\ldots,i_{\sigma (s)}\} $, where $s=|S|$ and $\sigma $ is a strictly increasing function from $\{ 1,\ldots,s\} $ to $\{ 1,\ldots,n\} $.  
For each $h=1,\ldots,n$, let $T_h=\{ i_1,\ldots,i_h\} $
and $T_0=\emptyset $.  Note that $S\cap T_{\sigma (\ell )}=\{ i_{\sigma (1)},\ldots,i_{\sigma (\ell )}\} $  and $m_{i_\sigma(\ell)} =v(T_{\sigma (\ell )} )-v(T_{\sigma (\ell )}\setminus \{ i_{\sigma (\ell )}\} )$ for all $\ell = 1, \ldots, s$.  
By $\sum _{i\in S}m_i<v(S)$, there exists some $\ell =1,\ldots,s$ such that 
\begin{align}\label{dec_mw}
v(T_{\sigma (\ell )} )-v(T_{\sigma (\ell )}\setminus \{ i_{\sigma (\ell )}\} )=m_{i_{\sigma (\ell )}}<v(S\cap T_{\sigma (\ell )})-v((S\cap T_{\sigma (\ell )})\setminus \{ i_{\sigma (\ell )}\} ).
\end{align}

Without loss of generality, we may assume that $\ell =s$ as follows.  Assume that $\ell <s$, but $\ell $ is the largest integer satisfying \eqref{dec_mw}.  
Then, %$%v(T_{\sigma (\ell ')} )-v(T_{\sigma (\ell ')}\setminus \{ i_{\sigma (\ell ')}\} )
 $m_{i_{\sigma (\ell ')}}\ge v(S\cap T_{\sigma (\ell ')})-v((S\cap T_{\sigma (\ell ')})\setminus \{ i_{\sigma (\ell ')}\} )$ for all 
$\ell '=\ell +1,\ldots,s$. Together with 
\[ %\sum _{\ell '=1}^s(v(T_{\sigma (\ell ')} )-v(T_{\sigma (\ell ')}\setminus \{ i_{\sigma (\ell ')}\} ))=
\sum _{\ell '=1}^sm_{i_{\sigma (\ell ')}}<v(S)=\sum _{\ell '=1}^s(v(S\cap T_{\sigma (\ell ')})-v((S\cap T_{\sigma (\ell ')})\setminus \{ i_{\sigma (\ell ')}\} )),\]
we have $\sum _{\ell '=1}^{\ell }m_{i_{\sigma (\ell ')}}<v(S\cap T_{\sigma (\ell )})$.  
Then, we can replace $S$ by $S\cap T_{\sigma (\ell )}$.  
Hereafter, we assume that $\ell =s$.  Note that $S\cap T_{\sigma (s)}=S$.  

By \eqref{dec_mw}, 
\begin{align}\label{nonconvex}
v(T_{\sigma (s)} )+v(S\setminus \{ i_{\sigma (s)}\} )<v(S)+v(T_{\sigma (s)}\setminus \{ i_{\sigma (s)}\} ).
\end{align}
%\begin{align*}
%v(\{ 1,...,i_{\sigma (\ell )}\} )-v(\{ 1,...,i_{\sigma (\ell )-1}\} )=m_{i_{\sigma (\ell )}}<v(\{ i_{\sigma (1)},...,i_{
%\end{align*}
%we have 
%\[ \sum _{j=1}^s(v(T(\sigma (j))-v(T(\sigma (j))\setminus \{ i_{\sigma (j)}\} ))<\sum _{j=1}^s(v(S\cap T(\sigma (j)))-v((S\cap T(\sigma (j)))\setminus \{ i_{\sigma (j)}\} ).\] 
We have 
$S\cup (T_{\sigma (s)}\setminus \{ i_{\sigma (s)}\} )=T_{\sigma (s)} $ and    
$S\cap (T_{\sigma (s)}\setminus \{ i_{\sigma (s)}\} )=S\setminus \{ i_{\sigma (s)}\} $ by $S\subseteq T_{\sigma (s)}$ and $i_{\sigma (s)}\in S$.   Since $v$ is 3-distant convex and $|S\setminus (T_{\sigma (s)}\setminus \{ i_{\sigma (s)}\} )|=|\{ i_{\sigma (s)}\} |=1$, we have that 
\begin{align}\label{dis1}
|(T_{\sigma (s)}\setminus \{ i_{\sigma (s)}\} )\setminus S|\le 1.
\end{align}
Moreover, \eqref{dis1} holds in equality; otherwise, \eqref{nonconvex} never holds because $T_{\sigma (s)}\setminus \{ i_{\sigma (s)}\} \subseteq  S$ implies 
$S = T_{\sigma(s)}$ by $S\subseteq T_{\sigma (s)}$ and $i_{\sigma(s)} \in S$.
%$S\cup (T_{\sigma (s)}\setminus \{ i_{\sigma (s)}\} )=T_{\sigma (s)}$ and $S\cap (T_{\sigma (s)}\setminus \{ i_{\sigma (s)}\} ) = S\setminus \{ i_{\sigma (s)}\} $.
Then, there exists some $h=1,\ldots,\sigma (s)-1$ such that $S=T_{\sigma (s)}\setminus \{ i_h\} $.  Therefore, $\sigma (s)=s+1$ and $S=\{ i_1,\ldots,i_{s+1}\} \setminus \{ i_h\} $.  Then,  
\begin{align*}
\sum _{i\in S}m_i&=\sum _{\ell =1}^{h-1}(v(T_{\ell })-v(T_{\ell -1}))+\sum _{\ell =h+1}^{s+1}(v(T_{\ell })-v(T_{\ell -1}))\text{ and } \\ 
v(S)&=\sum _{\ell =1}^{h-1}(v(T_{\ell })-v(T_{\ell -1}))+\sum _{\ell =h+1}^{s+1}(v(T_{\ell }\setminus \{ i_h\} )-v(T_{\ell -1}\setminus \{ i_h\} )),
\end{align*}
By $\sum _{i\in S}m_i<v(S)$, 
\begin{align*}%\label{ineq1}
\sum _{\ell =h+1}^{s+1}(v(T_{\ell })-v(T_{\ell -1}))<\sum _{\ell =h+1}^{s+1}(v(T_{\ell }\setminus \{ i_h\} )-v(T_{\ell -1}\setminus \{ i_h\} )).
\end{align*}
The left-hand side turns out to be 
\begin{align*}%\label{eq1}
\sum _{\ell =h+1}^{s+1}(v(T_{\ell })-v(T_{\ell -1}))&=\sum _{\ell =1}^{s+1}(v(T_{\ell })-v(T_{\ell -1}))-\sum _{\ell =1}^h(v(T_{\ell })-v(T_{\ell -1}))=v(T_{s+1})-v(T_{h}),
\end{align*}
and the right-hand side turns out to be 
\begin{align*}
\sum _{\ell =h+1}^{s+1}&(v(T_{\ell }\setminus \{ i_h\} )-v(T_{\ell -1}\setminus \{ i_h\} ))\\ 
&=\left(\sum _{\ell =1}^{h-1}(v(T_{\ell })-v(T_{\ell -1}))+\sum _{\ell =h+1}^{s+1}(v(T_{\ell }\setminus \{ i_h\} )-v(T_{\ell -1}\setminus \{ i_h\} )\right)-\sum _{\ell =1}^{h-1}(v(T_{\ell })-v(T_{\ell -1})) \\
&=v(S)-v(T_{h-1}).  
\end{align*}
%\begin{align*}
%v(T_{s})-v(T_{h})&=\sum _{\ell =h+1}^s(v(T_{\ell })-v(T_{\ell -1})) \\ 
%&<\sum _{\ell =h+1}^s(v(T_{\ell }\setminus \{ i_h\} )-v(T_{\ell -1}\setminus \{ i_h\} ))=v(S)-v(T_{h-1}),
%\end{align*}
Thus, $v(T_{s+1})+v(T_{h-1})<v(S)+v(T_{h})$.  By $S=T_{s+1}\setminus \{ i_h\} $, we have 
$T_{s+1}=S\cup T_h$ and $T_{h-1}=S\cap T_h$.  By 3-distant convexity of $v$ and $|T_h\setminus S|=|\{ i_h\} | = 1$, we have $|S\setminus T_h| \le 1$.  
By $S= T_{s+1} \setminus \{i_h\}$
and $h < s+1$,   $i_{s+1} \in S\setminus T_h$ and thus $S \setminus T_h = \{i_{s+1} \}$.
Therefore, $h=s$.  
We obtain the desired result by letting $k=s-1$, $k+1=s$, and $k+2=s+1$ by $S=\{ i_1,\ldots,i_{s-1},i_{s+1}\} $ and $T_{s}=\{ i_1,\ldots,i_{s-1}, i_s\} $.    
\hfill $\square $

\bigskip
 
% We say that $(N, v)$ is a \emph{2/3-convex} game if for any distinct $S_1, S_2, S_3$, there exist two distinct pairs $(S_i, S_j) \neq (S_k, S_l)$ ($i, j, k, l \in \{1, 2, 3\}$) such that $v(S_i \cup S_j) + v(S_i \cap S_j) \ge v(S_i) + v(S_j)$ and $v(S_k \cup S_l) + v(S_k \cap S_l) \ge v(S_k) + v(S_l)$. This concept was originally introduced by (\citet{MizutaniYoshida2024, MizutaniYoshida2026TwoThirdsSubmodular}) in the context of combinatorial optimization.
% It is known that every 2/3-convex game
% is $3$-distant convex \citep{Mizutani2024KDist}.

% Moreover, we show that every 2/3-convex game is permutationally convex, 
% a relaxed notion of convexity for TU games proposed by  \cite{Granot1982Perm} in  Appendix.
% {\color{orange} Using Theorem \ref{thm:mw}, there exists an ordering of players such that the corresponding marginal worth vector is in the core in a 2/3-convex game satisfying a mild condition.  }

% \begin{proposition}\label{claim_nonconvex}
% In a 2/3-convex game $(N,v)$ with $n\ge 3$, there exists a marginal worth vector that is in the core.
% Therefore, the core is nonempty.
% %Let $m$ be a marginal worth vector for some order of players $i_1,\ldots,i_n$.  
% %If $m$ is not in the core, then there exists some $k=1,\ldots,n-2$ such that 
% %\begin{align}\label{k+1k+2}
% %v(\{ i_1,\ldots,i_k,i_{k+1})+v(\{ i_1,\ldots,i_k,i_{k+2}\} )>v(\{ i_1,\ldots,i_k,i_{k+1},i_{k+2}\} )+v(\{ i_1,\ldots,i_k\} ).
% %\end{align}    
% \end{proposition}

We now show Theorem \ref{thm:core}.
The proof is constructive and yields an algorithm that finds a marginal worth vector in the core using \(O(n)\) evaluations of the characteristic function and \(O(n)\) arithmetic operations.

\bigskip
\noindent
{\bf Proof of Theorem \ref{thm:core}.}  By Lemma \ref{lem:2/3conveximplies3convex} and \ref{thm:mw}, it is sufficient to construct an ordering of players in which \eqref{k+1k+2} does not happen. 
Fix an arbitrary $i_n \in N$.
%If $n=2$, then the proof is done by the assumption that $v(\{ i^*\} )+v(\{ i\} )\le v(\{ i^*,i\} )$ where $N=\{ i^*,i\} $.  Thus, assume that $n\ge 3$.  
%We construct an ordering of players in which \eqref{k+1k+2} does not happen.  

%Fix an arbitrary $i_n\in N\setminus \{ i^*\} $.  

We begin with the case where $n\ge 5$.  Then, 
we can pick distinct $j,k\in N\setminus \{ i_n \} $.  
By  2/3-convexity of $v$, 
$v(N)+v(N\setminus \{ i_n,j\} )\ge v(N\setminus \{ i_n\} )+v(N\setminus \{ j\} )$ or 
$v(N)+v(N\setminus \{ i_n,k\} )\ge v(N\setminus \{ i_n\} )+v(N\setminus \{ k\} )$.    
Define $i_{n-1}=j$ if the former holds; otherwise define $i_{n-1}=k$.  Then, \eqref{k+1k+2} does not hold for $k=n-2$ regardless of the alignments of the players in $N\setminus \{ i_{n-1},i_n\} $.  

Let  $h=4,\ldots,n-1$.  Assume that $i_{h},\ldots,i_n\in N$ have already been chosen.  Let $S_h=N\setminus \{ i_{h+1},\ldots,i_n\} $.  Note that  $i_h\in S_h$ and  $|S_h| = h \ge 4$.  By $|S_h|\ge 4$, we can pick distinct  $j_h,k_h\in S_h \setminus \{ i_h\} $.  By 2/3-convexity of $v$,  $v(S_h)+v(S_h\setminus \{ i_h,j_h\} )\ge v(S_h\setminus \{ i_h\} )+v(S_h\setminus \{ j_h\} )$ or 
$v(S_h)+v(S_h\setminus \{ i_h,k_h\} )\ge v(S_h\setminus \{ i_h\} )+v(S_h\setminus \{ k_h\} )$.    
When $h\ge 5$, define $i_{h-1}=j_h$ if the former holds; otherwise define $i_{h-1}=k_h$.  Then, \eqref{k+1k+2} does not hold for $k=h-2$ with $h\ge 5$ regardless of the alignments of the players in $N\setminus \{ i_{h-1},\ldots,i_n\} $. 

Assume that $h=4$.  We need to define $i_{3}$ more carefully in this case.  Note that the rest of the proof applies to the case where $n \le 4$.   Let $\{ i,j,k\} =S_h\setminus \{ i_4\} =N\setminus \{ i_4,\ldots,i_{n} \} $.  
By 2/3-convexity of $v$, at least two of the following inequalities hold:   
\begin{align}
v(\{ i,j\} )&\ge v(\{ i\} )+v(\{ j\} ) \label{23a}\\ 
v(\{ j,k\})&\ge v(\{ j\} )+v(\{ k\} ) \label{23b} \\ 
v(\{ i,k\} )&\ge v(\{ i\} )+v(\{ k\} ) \label{23c}.
\end{align}
Without loss of generality, we may assume that \eqref{23a} and \eqref{23b} hold.  

By 2/3-convexity of $v$, 
%at least one of 
$v(S_4)+v(S_4\setminus \{ i_4,k\} )\ge v(S_4\setminus \{ i_4\} )+v(S_4\setminus \{ k\} )$ or 
$v(S_4)+v(S_4\setminus \{ i_4,i\} )\ge v(S_4\setminus \{ i_4\} )+v(S_4\setminus \{ i\} )$.  
Define $i_{3}=k$ if the former holds; otherwise define $i_{3}=i$. Then, \eqref{k+1k+2} does not hold for $k=2$. %with $h\ge 5$ regardless of the alignments $i_1$ and $i_2$.  %of the players in $N\setminus \{ i_{31},\ldots,i_n\} $. 

We consider the case where $i_3=k$.  The case where $i_3=i$ can be proved similarly.  
By 2/3-convexity of $v$, at least one of $v(\{ i,j,i_3\} )+v(\{ i\} )\ge v(\{ i,i_3\} )+v(\{ i,j\} )$ or $v(\{ i,j,i_3\} )+v(\{ j\} )\ge v(\{ j,i_3\} )+v(\{ i,j\} )$ holds.  Define $i_2=j$ if the former holds; otherwise define$i_2=i$.  
Finally, let the remaining player be $i_1$. Then, $v(\{ i_1,i_2\} )\ge v(\{ i_1\} )+v(\{ i_2\} )$ since $\{ i_1,i_2\} =\{ i,j\} $ and \eqref{23a} holds.\footnote{In the case where $i_3=i$, $\{ i_1,i_2\} =\{ j,k\} $ and $v(\{ i_1,i_2\} )\ge v(\{ i_1\} )+v(\{ i_2\} )$ is guaranteed by \eqref{23b}.}   
Hence,  
\eqref{k+1k+2} never holds for any successive pair in $i_1,\ldots,i_n$ by construction.
By Lemma \ref{thm:mw}, the marginal worth vector corresponding to $i_1,\ldots,i_n$ is in the core.  
\hfill $\square $

\subsection{ Relationship  to permutational convexity} 

Here, we denote  $[k] = \{1, \cdots, k\}$ for $k \ge 1$
and $[0]=\emptyset$.
A TU game $(N,v)$ is called \textit{permutationally convex} if there exists a labeling $1,2,\ldots,n$ of the players such that 
\begin{align}
\label{eq:perm_submo}
v([j]\cup S)+v([k])\le v([k]\cup S)+v([j])
\end{align}
for all $S\subseteq N\setminus [k]$ and $k\ge j\ge 0$.

\cite{Granot1982Perm} introduced this concept and showed that
the marginal worth vector corresponding to the labeling is in the core.
One might expect that every $2/3$-covex game is permutationally convex.
However, the following example shows that this is not true.

% While Proposition \ref{prop:23-conv} shows that a $2/3$-convex game with the superadditivity condition for singletons is permutationally convex, the following example shows that in general, a $2/3$-convex game is not necessarily permutationally convex.

\begin{example}
\label{ex:non-perm}
Let $K_n=(V,E)$ be a \emph{complete graph} on $n$ vertices, i.e., every pair of distinct vertices in $V$ is connected by a unique edge in $E$.
Take any pair of vertices $v_1,v_2\in V$.
We define the weight function $w:E\to \mathbb{Z}$ as follows:
\begin{align*}
w(e):=\begin{cases}
    1 & \mathrm{if\ }e=v_1v_2,\\
    -1 & \mathrm{otherwise.}
\end{cases}
\end{align*}
Define the cut function $f:2^V\to \mathbb{Z}$ of $G$ as follows:
\begin{align*}
f(X):=\sum_{u\in X,v\in V\setminus X}w(uv) = 
\begin{cases}
- |X| | V \setminus  X| + 2 & \text{ if } |X \cap \{v_1, v_2\}| = 1 \\
- |X| | V \setminus  X|     & \text{ otherwise}.
\end{cases}
\end{align*}
Then, $f$ is 2/3-convex.
Indeed, the supermodular inequality of $f$ does not hold only for the pairs of $X\cup \{v_1\}$ and $X\cup \{v_2\}$ for every $X\subseteq V\setminus \{v_1,v_2\}$.
To see this, define $g(X) = -|X||V \setminus X|$ for all $X \subseteq V$.
Then, $g$ is convex by $g(S \cup T) + g(S \cap T) - ( g(S) + g(T) ) = 2 |S \setminus T| |T \setminus S|$.
Let $S, T \subseteq V$.
If $|S \cap \{v_1, v_2\}| \neq 1$ or $|T \cap \{v_1, v_2\}| \neq 1$,
then the supermodular inequality of $f$ holds by convexity of $g$
and the definition of $f$.
Thus, we assume that $|S \cap \{v_1, v_2\}| = 1$ and $|T \cap \{v_1, v_2\}| = 1$.
Without loss of generality, we may assume that $v_1 \in S$ and $v_2 \notin S$.
If $v_1 \in T$ and $v_2 \notin T$,  then the supermodular inequality of $f$ holds by convexity of $g$
and the definition of $f$.
If $v_2 \in T$ and $v_1 \notin T$, then
$$
f(S\cup T) + f(S\cap T) - (f(S) + f(T)) = 2|S \setminus T| |T \setminus S|  - 4.
$$
This is negative if only if $|S \setminus T| = 1$ and $|T \setminus S| = 1$
since $v_1 \in S \setminus T$ and $v_2 \in T \setminus S$.

We show that $f$ is not permutationally convex.
Take any labeling $1,2,\ldots,n$ of vertices in $V$.
Let $i_1$ and $i_2$ be the labels of $v_1$ and $v_2$, respectively.
Without loss of generality, we may assume that $i_1< i_2$.
Then, $f$ does not satisfy (\ref{eq:perm_submo}) for $S=\{i_2\}$,  $k=i_1$, and $j=i_1-1$, because the supermodular inequality of $f$ for $[i_1-1]\cup \{i_2\}$ and $[i_1]=[i_1-1]\cup \{i_1\}$ does not hold by
\begin{align*}
f(\{1, \ldots, i_1 -1, i_2\} ) - f(\{1, \ldots, i_1 -1\} ) &= -|V| + 2i_1 +1 \\
&> -|V| + 2i_1 -1 \\
&= f(\{1, \ldots, i_1, i_2\} ) - f(\{1, \ldots, i_1 \} ).
\end{align*}
Thus, $f$ is not permutationally convex.
\end{example}

We next show that a $2/3$-convex game with superadditivity condition for singletons is permutationally convex.
%We show the following relation between 2/3-convex games and permutationally convex games.
\begin{proposition}
\label{prop:23-conv}
Let $(N,v)$ be a 2/3-convex game such that $v(\{i\})+v(\{j\})\le v(\{i,j\})$ for all $i,j\in N$ with $i\ne j$.
Then, $(N,v)$ is permutationally convex.
\end{proposition}

Before presenting the proof, we note a computational implication of the proposition. The definition of permutational convexity does not itself provide a method for finding a labeling that satisfies \eqref{eq:perm_submo}. The proof below gives a constructive procedure for finding such a labeling using \(O(n^3)\) evaluations of the characteristic function and \(O(n^3)\) arithmetic operations.

\bigskip
\noindent
{\bf Proof of Proposition \ref{prop:23-conv}.}
Since $(N,v)$ is 2/3-convex, it is also 3-distant convex.
Hence, for any labeling $1,2,\ldots,n$ of the players, (\ref{eq:perm_submo}) holds if $\left|([j]\cup S)\bigtriangleup [k]\right|=k-j+|S|\ge 3$, which implies that (\ref{eq:perm_submo}) holds if $|S|\ge 2$ or $k-j\ge 2$.
Thus, it suffices to construct a labeling $1,2,\ldots,n$ that satisfies (\ref{eq:perm_submo}) for all $S\subseteq N\setminus [k]$ with $|S|=1$, and all $k\ge 1$ and $j=k-1$, that is,
\begin{align}
\label{ineq:label_condition}
v([k-1]\cup \{\ell\})+v([k])\le v([k]\cup \{\ell\})+v([k-1])
\end{align}
for all $\ell\in N\setminus [k]$ and $k\ge 1$.

We inductively construct a labeling $1,2,\ldots,n$ that satisfies (\ref{ineq:label_condition}) for all $\ell\in N\setminus [k]$ and $k\ge 1$.
We first choose any player and assign the label $1$ to it.
Then, (\ref{ineq:label_condition}) holds for all $\ell\in N\setminus [1]$ and $k=1$ because $v(\{i\})+v(\{j\})\le v(\{i,j\})$ for all $i,j\in N$ with $i\ne j$.
Assume that we have a labeling $1,2,\ldots,i$ such that (\ref{ineq:label_condition}) holds for all $\ell\in N\setminus [k]$ and $1 \le k\le i$.
By induction, it suffices to construct another labeling $1,2,\ldots,i+1$ of size $i+1$ such that (\ref{ineq:label_condition}) holds for all $\ell\in N\setminus [k]$ and $1 \le k\le i+1$.

Choose any player in $N\setminus [i]$ and assign the label $i+1$ to it.
The following claim completes the proof by induction.
Note that the assumption in the following claim holds for $p=i+1$ by induction hypothesis for the labeling $1,2,\ldots,i$.
\begin{claim}
\label{claim:perm_induction}
Let $p$ be a positive integer with $2\le p\le i+1$.
We define a labeling $1',2',\ldots,(i+1)'$ as follows:
\begin{align*}
j':=\begin{cases}
    j & \mathrm{if\ }1\le j\le p-1,\\
    i+1 & \mathrm{if\ }j=p,\\
    j-1 & \mathrm{if\ }p+1\le j\le i+1.
\end{cases}
\end{align*}
Suppose that the labeling $1',2',\ldots,(i+1)'$ satisfies (\ref{ineq:label_condition}) for all $\ell\in N\setminus [k']$ and $1 \le k\le i+1$ with $k\ne p$  where $[k']=\{1', 2', \ldots, k'\}$.
Then, one of the following conditions holds:
\begin{itemize}
    \item The labeling $1',2',\ldots,(i+1)'$ satisfies (\ref{ineq:label_condition}) for all $\ell\in N\setminus [k']$ and $k=p$.
    \item Let $1'',2'',\ldots,(i+1)''$ be another labeling obtained by exchanging $i+1$ and $p-1$ in the labeling $1',2',\ldots,(i+1)'$.
    If $p\ge 3$, then $1'',2'',\ldots,(i+1)''$ satisfies (\ref{ineq:label_condition}) for all $\ell\in N\setminus [k'']$ and $1 \le k\le i+1$ with $k\ne p-1$
    where
    $[k''] = \{1'', 2'', \ldots, k''\}$. If $p=2$, then $1'',2'',\ldots,(i+1)''$ satisfies (\ref{ineq:label_condition}) for all $\ell\in N\setminus [k'']$ and $1 \le k\le i+1$.
\end{itemize}
\end{claim}
\begin{proof}[Proof of Claim \ref{claim:perm_induction}]
%We will use the notation $[q']:=\{1',2',\ldots,q'\}$ and $[q'']:=\{1'',2'',\ldots,q''\}$ for convenience.
Suppose that the first condition does not hold.
Then, there exists $\ell'\in N\setminus [p']$ such that
\begin{align*}
v([(p-1)']\cup \{\ell'\})+v([p'])>v([p']\cup \{\ell'\})+v([(p-1)']).
\end{align*}
This implies that
\begin{align}
\label{ineq:induc_p}
&v([p-1]\cup \{\ell'\})+v([p-1]\cup \{i+1\})=v([(p-1)']\cup \{\ell'\})+v([p'])\notag\\&>v([p']\cup \{\ell'\})+v([(p-1)'])=v([p-1]\cup \{i+1,\ell'\})+v([p-1]).
\end{align}
Recall that $1'',2'',\ldots,(i+1)''$ is the labeling obtained by exchanging $i+1$ and $p-1$ in the labeling $1',2',\ldots,(i+1)'$, that is, $1,2,\ldots,p-2,i+1,p-1,\ldots,i$.
By (\ref{ineq:induc_p}) and 2/3-convexity of $v$, we have
\begin{align*}
&v([(p-1)'']\cup \{\ell\})+v([p''])=v([p-2]\cup \{i+1,\ell\})+v([p-1]\cup \{i+1\})\\&\le v([p-1]\cup \{i+1,\ell\})+v([p-2]\cup \{i+1\})=v([p'']\cup \{\ell\})+v([(p-1)''])
\end{align*}
for all $\ell\in N\setminus [p'']$.
Hence, the labeling $1'',2'',\ldots,(i+1)''$ satisfies (\ref{ineq:label_condition}) 
%for all  $\ell\in N\setminus [k]$ 
%and $k=p$ 
for all $\ell\in N\setminus [p'']$ and $k=p$.
Since $j''=j'$ for all $j=1,2,\ldots,p-2$, the labeling $1'',2'',\ldots,(i+1)''$ also satisfies (\ref{ineq:label_condition}) for all $\ell\in N\setminus [k'']$ and $1 \le k\le p-2$ by the assumption.
Moreover, since $[j'']=[j']$ for all $j=p,p+1,\ldots,i+1$, the labeling $1'',2'',\ldots,(i+1)''$ satisfies (\ref{ineq:label_condition}) for all $\ell\in N\setminus [k'']$ and $p+1\le k\le i+1$ by the assumption.
Therefore, $1'',2'',\ldots,(i+1)''$ satisfies (\ref{ineq:label_condition}) for all $\ell\in N\setminus [k'']$and $1 \le k\le i+1$ with $k\ne p-1$, as desired if $p\ge 3$.
If $p=2$, then since $v(\{i\})+v(\{j\})\le v(\{i,j\})$ for all $i,j\in N$ with $i\ne j$, $1'',2'',\ldots,(i+1)''$ also satisfies (\ref{ineq:label_condition}) for all $\ell\in N\setminus [1'']$ and $k=1$, as desired.  %and $k=1$, as desired.
\end{proof}
\hfill $\square $

% admits a marginal worth vector in the core, 
% but do not provide a method for computing such a vector.\footnote{Indeed, 
% the marginal worth vector corresponding to the labeling in the above definition is in the core.
% }
% In contrast, for $3$-distant convex games, such a vector can be computed 
% in polynomial time using the argument in Proposition \ref{prop:marginal}.
% }

\begin{remark}
The proof of Proposition \ref{prop:23-conv} gives the construction of a labeling of players that satisfies the condition of permutational convexity and yields a marginal worth vector in the core.
Recall that the proof of Theorem \ref{thm:core} also provides the construction of an ordering of players such that the corresponding marginal worth vector is in the core.
We give an example showing that such an ordering in the proof of Theorem \ref{thm:core} need not satisfy the condition (\ref{eq:perm_submo}) of permutational convexity.

Let $V=\{v_1,v_2,v_3,v_4\}$.
As in Example \ref{ex:non-perm}, let $K_4=(V,E)$ be a complete graph on $4$ vertices.
Define the weight function $w:E\to \mathbb{Z}$ as follows:
\begin{align*}
w(e):=\begin{cases}
    1 & \mathrm{if\ }e=v_2v_4,\\
    -1 & \mathrm{otherwise.}
\end{cases}
\end{align*}
Let $f:2^V\to \mathbb{Z}$ be the cut function defined as in Example \ref{ex:non-perm}.
Define $g:2^V\to \mathbb{Z}$ as follows:
\begin{align*}
g(X):=\begin{cases}
    f(X)-2 & \mathrm{if\ }X\ne \emptyset,\\
    f(X) & \mathrm{if\ }X= \emptyset.
\end{cases}
\end{align*}
By Example \ref{ex:non-perm}, the supermodular inequality of $g$ does not hold only for the pairs of $X\cup \{v_2\}$ and $X\cup \{v_4\}$ for $X=\{v_1\},\{v_3\},\{v_1,v_3\}$.
Then, $g$ is $2/3$-convex and satisfies superadditivity for every pair of singletons.
Since $g$ satisfies 
\begin{align*}
&g(\{v_1,v_2,v_3,v_4\})+g(\{v_1,v_2\})\ge g(\{v_1,v_2,v_3\})+g(\{v_1,v_2,v_4\})\ \mathrm{and}\\
&g(\{v_1,v_2,v_3\})+g(\{v_1\})\ge g(\{v_1,v_3\})+g(\{v_1,v_2\}),
\end{align*}
the construction in the proof of Theorem \ref{thm:core} may produce the ordering $v_1,v_2,v_3,v_4$.
However, the above ordering does not satisfy the condition (\ref{eq:perm_submo}) of permutational convexity because
\begin{align*}
g(\{v_1,v_4\})+g(\{v_1,v_2\})> g(\{v_1,v_2,v_4\})+g(\{v_1\}).
\end{align*}
\end{remark}

\section{Application}

Consensus-based decision-making gives each participant an effective
veto. The OSCE provides a ``consensus minus two'' exception under which
neither of two States involved in a dispute can veto a decision
directing them to seek conciliation
\citep[Annex~4, paras.~1 and 4]{CSCE1992Stockholm};
see also \citet{OSCEDecisionMaking}.

%Motivated by this procedure, we consider a decision involving two
%designated disputing participants \(e,f\in N\). Let \(S\subseteq N\)
%denote the set of participants that do not object to the decision. We
%represent the value generated by the decision by
Motivated by this procedure, we consider a decision involving two
designated disputing participants \(e,f\in N\). Let \(S\subseteq N\)
denote the set of participants that do not object to the decision.
The characteristic function represents the value generated by the
adopted decision, rather than merely indicating whether it is adopted.
Although neither disputing participant can veto the decision, the
value generated by it may depend on whether one or both of them object,
because conciliation may be less effective without their support.
We therefore define
\[
v(S)=
\begin{cases}
\alpha, & S=N,\\
\beta_e, & S=N\setminus\{e\},\\
\beta_f, & S=N\setminus\{f\},\\
\gamma, & S=N\setminus\{e,f\},\\
0, & \text{otherwise}.
\end{cases}
\]

Here, \(\alpha\) is the value generated when no participant objects,
\(\beta_e\) and \(\beta_f\) are the values generated when only \(e\)
or only \(f\) objects, respectively, and \(\gamma\) is the value
generated when both disputing participants object. We assume that
$ \alpha\geq\max\{\beta_e,\beta_f\}$ and
$\min\{\beta_e,\beta_f\}\geq\gamma\geq 0$.
% Thus, an objection by either disputing participant does not prevent
% adoption, although it may reduce the value generated by the decision.
% An objection by any other participant prevents adoption.
This assumption guarantees that $v$ is monotone, that is, $v(S) \le v(T)$ for all $S, T \subseteq N$
with $S \subseteq T$.
Then, $v$ need not be convex. For example, letting
$\alpha=5,
\beta_e=\beta_f=4,
\gamma=1, $  we have
$$
v(N)+v(N \setminus \{e, f\}) -v(N \setminus \{e\} )-v(N \setminus \{f\} )
=5+1-4-4 = -2.
$$

\begin{proposition}
$v$ is 2/3-convex.
\end{proposition}
\begin{proof}
By the definition of  2/3-convexity,
it is sufficient to show
that for any $S, T \subseteq N$,
if $v(S \cup T) + v(S \cap T) < v(S) + v(T)$,
then $S, T \in \{N \setminus \{e\}, N \setminus \{f\} \}$.

Take any $S, T \subseteq N$.
Suppose, for contradiction, that $v(S \cup T) + v(S \cap T) < v(S) + v(T)$
and $S \notin \{ N \setminus \{e\}, N \setminus \{f\} \}$.
Note that $S \setminus T \neq \emptyset$ and $T \setminus S \neq \emptyset$ hold.
There are two cases to consider.

\bigskip

\noindent{Case 1.} 
 We assume that $S= N \setminus \{e, f\}$.
%By $T \setminus S \neq \emptyset$, $T \cap \{e, f\} \neq \emptyset $
%and thus $v(S \cup T) \ge \min \{\beta_e,\beta_f\} $.
By $S \setminus T \neq \emptyset$, $T$ does not contain some player in $N \setminus \{e, f\}$.
Thus, $v(T) = 0$ and $v(S \cap T) = 0$.
By monotonicity of $v$,
$
v(S \cup T) + v(S \cap T) = v(S \cup T) \ge  v(S) = v(S)+ v(T),
$
a contradiction.

\bigskip
\noindent{Case 2.} 
We assume that $S \neq N \setminus \{e, f\}$.
By $S \notin \{ N \setminus \{e\}, N \setminus \{f\}, N \}$,
$S$ does not contain some player in $N \setminus \{e, f\}$.
Thus, $v(S) = v(S \cap T) = 0$.
By monotonicity of $v$,
$
v(S \cup T) + v(S \cap T) = v(S \cup T) \ge  v(T) = v(S)+ v(T),
$
a contradiction.
\end{proof}

\section{Conclusion}

This paper has studied the core of \(2/3\)-convex games. We have shown
that every \(2/3\)-convex game admits a marginal worth vector in the
core and therefore has a nonempty core. 
We
have also clarified the relationship between 2/3-convexity
and permutational convexity.

An important direction for future research is to examine other solution
concepts for \(2/3\)-convex games. Although a marginal worth vector is
always in the core, the Shapley value need not be. It would therefore
be useful to identify additional conditions under which the Shapley
value belongs to the core.  There are many other important properties of various solution concepts in convex games.  See, for example, \citet{PelegSudholter2007CooperativeGames} and references therein. It would be interesting to investigate how \(2/3\)-convexity
affects these properties.  %other solutions, such as the nucleolus and the least core.

Every \(2/3\)-convex game is \(3\)-distant convex, but the latter class
is substantially broader and does not generally guarantee core
nonemptiness. In a companion paper \citep{BandoHiraiImamuraMizutani2026}, we study \(3\)-distant convex games
in detail, providing a necessary and sufficient condition for core
nonemptiness and a strongly polynomial-time combinatorial algorithm
for constructing a core allocation whenever one exists.
Together, these two papers demonstrate that these new notions of convexity provide a tractable domain for cooperative-game analysis and capture economically and institutionally relevant applications beyond convex games.
% While Proposition \ref{prop:23-conv} shows that a $2/3$-convex game with the superadditivity condition for singletons is permutationally convex, the following example shows that in general, a $2/3$-convex game is not necessarily permutationally convex.

% \begin{example}
% \label{ex:non-perm}
% Let $K_n=(V,E)$ be a \emph{complete graph} on $n$ vertices, i.e., every pair of distinct vertices in $V$ is connected by a unique edge in $E$.
% Take any pair of vertices $v_1,v_2\in V$.
% We define the weight function $w:E\to \mathbb{Z}$ as follows:
% \begin{align*}
% w(e):=\begin{cases}
%     1 & \mathrm{if\ }e=v_1v_2,\\
%     -1 & \mathrm{otherwise.}
% \end{cases}
% \end{align*}
% Define the cut function $f:2^V\to \mathbb{Z}$ of $G$ as follows:
% \begin{align*}
% f(X):=\sum_{u\in X,v\in V\setminus X}w(uv).
% \end{align*}
% Then, $f$ is 2/3-convex.
% Indeed, the supermodular inequality of $f$ does not hold only for the pairs of $X\cup \{v_1\}$ and $X\cup \{v_2\}$ for every $X\subseteq V\setminus \{v_1,v_2\}$.

% We show that $f$ is not permutationally convex.
% Take any labeling $1,2,\ldots,n$ of vertices in $V$.
% Let $i_1$ and $i_2$ be the labels of $v_1$ and $v_2$, respectively.
% Without loss of generality, we may assume that $i_1< i_2$.
% Then, $f$ does not satisfy (\ref{eq:perm_submo}) for $S=\{i_2\}$,  $k=i_1$, and $j=i_1-1$, because the supermodular inequality of $f$ for $[i_1-1]\cup \{i_2\}$ and $[i_1]=[i_1-1]\cup \{i_1\}$ does not hold.
% Thus, $f$ is not permutationally convex.
% \end{example}

\section*{AI disclosure}
We used OpenAI's GPT models and Anthropic's Claude models to assist in  improving the presentation of the manuscript.
%In particular, the proofs of Propositions \ref{prop:symmetric_oligopoly} and \ref{prop:asymmetry_oligopoly} were obtained with the assistance of ChatGPT.
The authors take full responsibility for the content of this paper, including all mathematical claims and proofs.

\bibliographystyle{chicago}
\bibliography{reference}

\end{document}